\documentclass[fleqn,twoside]{article}
\usepackage{espcrc2}

\usepackage{graphicx}
\usepackage[figuresright]{rotating}

\newcommand{\AmS}{{\protect\the\textfont2
  A\kern-.1667em\lower.5ex\hbox{M}\kern-.125emS}}

\title{Measurement of $|V_{ub}|$ in semi-inclusive charmless $B \to \pi X$ decays }

\author{C. S. Kim\address[MCSD]{Department of Physics and IPAP, Yonsei University, Seoul,
120-749, Korea }%
        \thanks{The work of C.S.K. was supported
               in part by  CHEP-SRC Program, Grant No. 20015-111-02-2
               and Grant No. R03-2001-00010 of the KOSEF,
               in part by BK21 Program and Grant No. 2001-042-D00022 of the KRF,
               and in part by Yonsei Research Fund, Project No. 2001-1-0057.},
        Jake Lee\addressmark\thanks{The work of J.L. was supported by BK21 Program
                 and Grant No. 2001-042-D00022 of the KRF.}
        and
        Sechul Oh\addressmark\thanks{The work of S.O. was supported by BK21 Program
                 and Grant No. 2001-042-D00022 of the KRF.} }
\begin{document}

\begin{abstract}
We study semi-inclusive charmless decays $B \to \pi X$,
where $X$ does not contain a charm (anti)quark.
The mode $\bar B^0 \to \pi^- X$ turns out to be be particularly
useful for determination of the CKM matrix element $|V_{ub}|$.
We present the branching ratio (BR) of $\bar B^0 \to \pi^- X$
as a function of $|V_{ub}|$, with an estimation of possible
uncertainty.
The BR is expected to be an order of $10^{-4}$.
\vspace{1pc}
\end{abstract}

\maketitle

\section{Introduction}

A precise measurement of the Cabibbo-Kobayashi-Maskawa (CKM) matrix elements
\cite{ckm} is one of the key issues
in the study of $B$ mesons and $B$ factory experiments.
In particular, the accurate determination of $V_{ub}$ is one of the most
challenging problems in $B$ physics.

Theoretical and experimental studies for probing $V_{ub}$ have been mostly
focused on the semileptonic $B$ meson decays.  The CLEO
result obtained using the exclusive semileptonic decay
$B \to \rho l \bar \nu$ \cite{cleo} : (in $10^{-3}$)
\begin{eqnarray}
  |V_{ub}| = ( 3.25 \pm 0.14^{+0.21}_{-0.29}
  \pm 0.55 (\rm model) ).
\end{eqnarray}
The OPAL data obtained using the inclusive decay $B \to X_u l \bar \nu$
\cite{opal} : (in $10^{-3}$)
\begin{eqnarray}
  |V_{ub}| = ( 4.00 \pm 0.65^{+0.67}_{-0.76}
  \pm 0.19 (\rm HQE) ).
\end{eqnarray}

In this work \cite{klo} we study semi-inclusive charmless decays $B \to \pi X$ and
investigate the possibility of extracting $|V_{ub}|$ from these processes.
Compared to the exclusive decays, these semi-inclusive decays are generally
expected to have less hadronic uncertainty and larger branching ratios.
There are several possible processes in $B \to \pi X$ type decays, such as
$\bar B^0 \to \pi^{\pm (0)} X$, $B^0 \to \pi^{\pm (0)} X$,
$B^{\pm} \to \pi^{\pm (0)} X$, where $X$ does not contain a charm
(anti)quark.
Among these processes of the type $B \to \pi X$, we identify a certain mode,
$\bar B^0 \to \pi^- X$, whose analysis is theoretically clean and which
can be used for determining $|V_{ub}|$.  Then, we calculate the branching ratio
(BR) of $\bar B^0 \to \pi^- X$, and present the result as a function of
$|V_{ub}|$ with an estimation of possible uncertainty.   We also consider
the $B^0 - \bar B^0$ mixing effect through $\bar B^0 \to B^0 \to \pi^- X$.

\section{Classification of semi-inclusive charmless $B \to \pi X$ decays}

Among the semi-inclusive charmless $B \to \pi X$ decays, let us first consider
the mode $\bar B^0 \to \pi^- X$.  Contributions for the decay amplitude of
this mode arise from the color-favored tree ($b \to u \bar u d$) diagram and
the $b \to d$ penguin diagram, so that the tree diagram contribution
dominates.
The charged pion $\pi^-$ in the final state can be produced
via a $W$ boson emission at tree level and is expected to be energetic
($\sim m_B / 2$).  The decay amplitude can be written as
\begin{eqnarray}
&\mbox{}& A( \bar B^0 \to \pi^- X ) \nonumber \\
&=& A (b \to \pi^- u) \cdot h(u \bar d \to X(u \bar d))~,
\end{eqnarray}
where $h$ denotes a hadronization function describing the combination of
the $u \bar d$ pair to make the final state $X$.  To obtain the decay rate,
$X(u \bar d)$ should be summed over all the possible states, such as $\pi^+ \pi^0$,
$\pi \pi \pi$ etc, so this process is effectively a \emph{two}-body
decay process of $b \to \pi^- u$.   Thus, in this mode, no hadronic form factors
are involved, and as a result the model-dependence and uncertainty relevant to
hadronic form factors do not appear.
We note that the \emph{energetic} charged pion $\pi^-$ in the final
state can be a characteristic signal for this mode.
(The net electric charge of $X$ should be \emph{positive} so that $\pi^-$
cannot be produced in the case of $X = \pi \pi$.)

Now let us consider the mode $B^- \to \pi^- X$.
Various contributions are responsible for this process :
the color-favored tree diagram, the color-suppressed tree diagram, the $b \to d$
and $b \to s$ penguin diagrams.  The color-favored tree contribution
and one of $b \to d$ penguin contributions are
similar to those in $\bar B^0 \to \pi^- X$, which are effectively two-body type
($b \to \pi^- u$) processes.  But, the color-suppressed tree and
other penguins differ from those in $\bar B^0 \to \pi^- X$.
In fact, these diagrams correspond to a three-body decay process
of $B^- \to \pi^- u \bar u$ in the parton model approximation.
The charged pion $\pi^-$ in the final state
contains the spectator antiquark $\bar u$. So the analysis involves
the hadronic form factor for the $B \to \pi$ transition which is model-dependent.
Furthermore, the $b \to s$ penguin contribution is not suppressed compared to
the tree contributions, but dominant in this mode.
Therefore, compared to the case of $\bar B^0 \to \pi^- X$,
the analysis of this mode is much more complicated and involves larger uncertainty.

Other modes of the type $B \to \pi X$ can be similarly classified.
For instance, in the mode $B^0 \to \pi^- X$, the color-favored tree
($\bar b \to \bar u u \bar d$ and $\bar b \to \bar u u \bar s$) diagrams and
the $b \to d$ and $b \to s$ penguin diagrams are responsible for the decay process.
In this case, the charged pion $\pi^-$ contains the spectator quark $d$ so that
the process is effectively a three-body decay $B^0 \to \pi^- u \bar d ~(\bar s)$
and the hadronic form factor for the $B \to \pi$ transition is involved.
Other processes are effectively
a combination of the two-body decay process ($b \to \pi q$) and the three-body
decay process ($B^- \to \pi^- q \bar q^{\prime}$).

\section{Analysis of $\bar B^0 \to \pi^- X$ decay}

We have seen that the process $\bar B^0 \to \pi^- X$ is
particularly interesting, because it is effectively the two-body decay
process $b \to \pi^- u$ and no uncertainty from hadronic form factors is
involved.   Thus, its theoretical analysis is expected to be quite clean.

We calculate the BR of the process
$\bar B^0 \to \pi^- X$, where $X$ can contain an up quark and a down antiquark.
We use the effective Hamiltonian and the effective Wilson coefficients
given in Ref. \cite{desh}.
The BR can be expressed as a polynomial of $|V_{ub}|$:
\begin{eqnarray}
&\mbox{}& {\cal B}(\bar B^0 \to \pi^- X)   \nonumber \\
    &=& \left|{V_{ub} \over 0.004} \right|^2 \cdot {\cal B}_2
    + \left|{V_{ub} \over 0.004}
    \right| \cdot {\cal B}_1 + {\cal B}_0 ~,
\label{BR}
\end{eqnarray}
where for convenience we have scaled $|V_{ub}|$ by the factor 0.004
(the central value of the OPAL data).

\includegraphics[width=20pc]{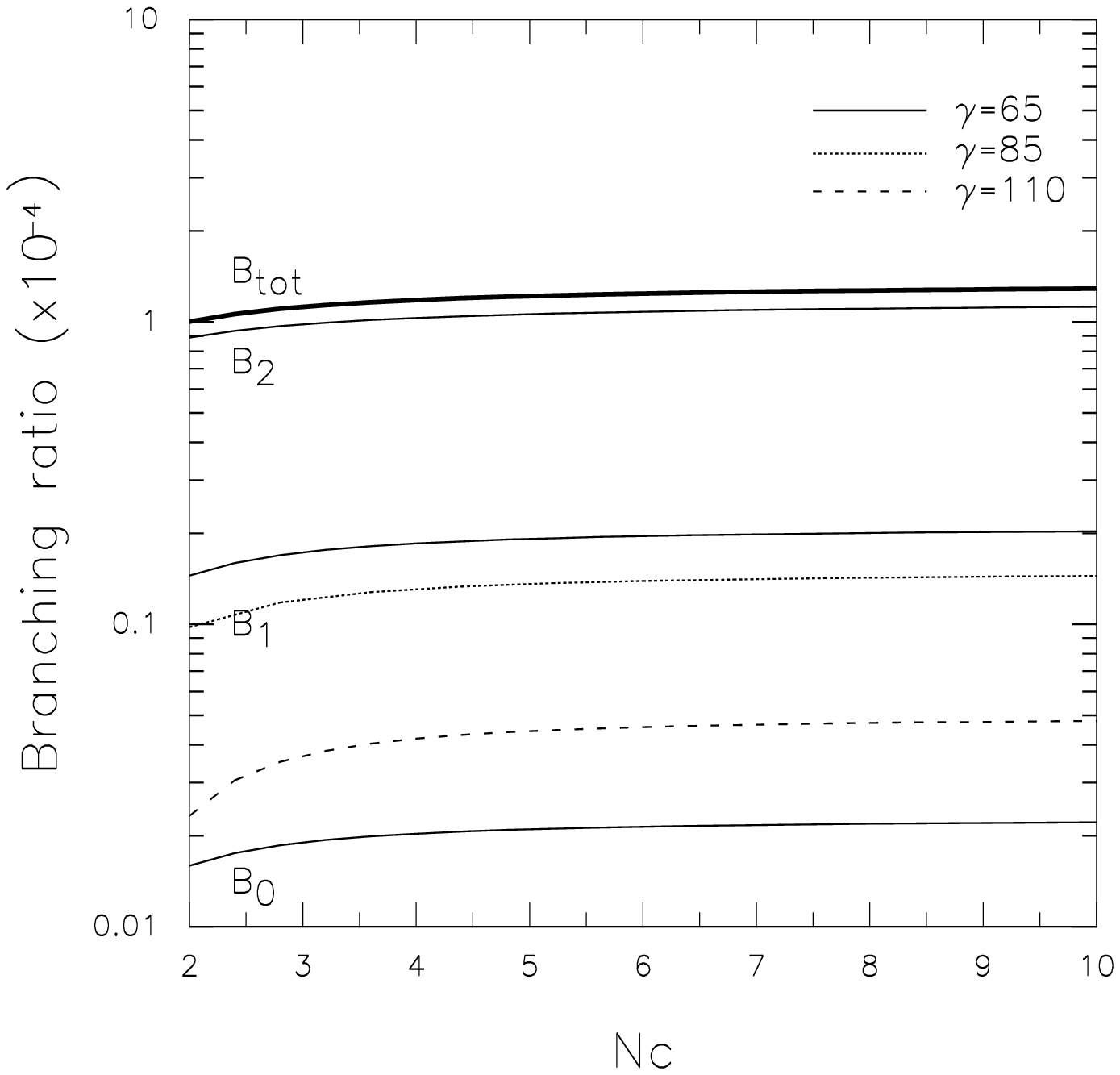}
\vspace{-3cm}
\\
Figure 1. The branching ratio (in $10^{-4}$) versus $N_c$
for $\bar B^0 \to \pi^- X$ decay. $B_{tot} (\equiv {\cal B})$ has been
calculated using $|V_{ub}| = 0.004$ and is denoted by the bold solid line.
The solid, dotted, and dashed lines
correspond to $\gamma = 65^0,~85^0,~110^0$, respectively. \\ \\

In Figure 1, we present the BR of $\bar B^0 \to \pi^- X$ as a function of
the effective number of color $N_c$ for three different values of
the CP phase angle $\gamma (\equiv \phi_3)= 65^0,~ 85^0,~ 110^0$.
${\cal B}_2$ and ${\cal B}_0$ are independent of $\gamma (\equiv \phi_3)$,
and only ${\cal B}_1$ depends on $\gamma (\equiv \phi_3)$.
Three different lines for ${\cal B}_1$ correspond to the relevant values
of $\gamma (\equiv \phi_3)$, respectively.
It is clearly shown that ${\cal B}_2$ is dominant.
An reperesentative value of ${\cal B}$ for
$|V_{ub}| =0.004$ and $\gamma (\equiv \phi_3)= 85^0$ is shown as the bold
solid line in the figure.
The value of ${\cal B}$ does not vary much as $N_c$ varies.

\includegraphics[width=20pc]{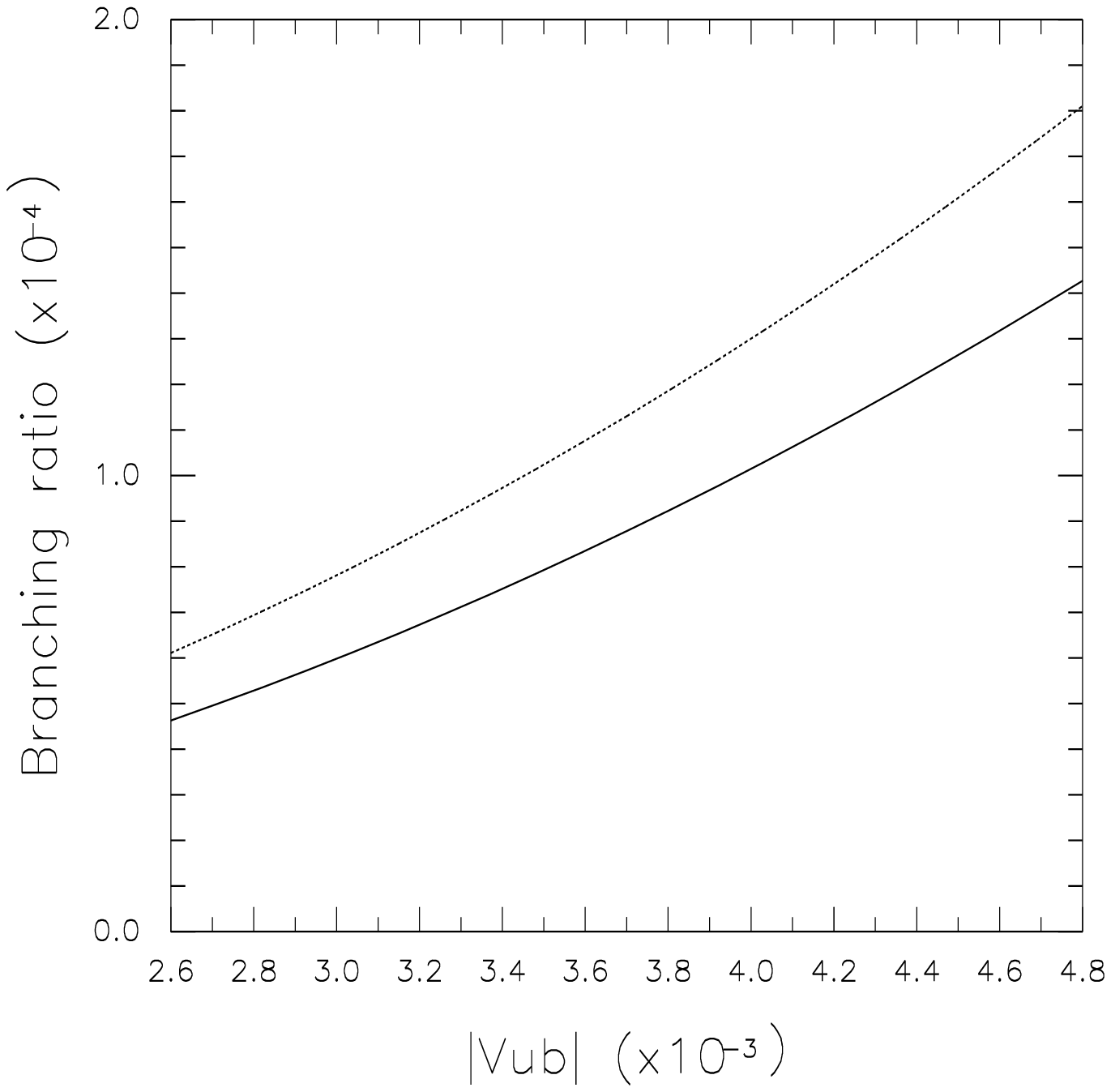}
\vspace{-3cm}
\\
Figure 2. The branching ratio (in $10^{-4}$) versus $|V_{ub}|$ for
$\bar B^0 \to \pi^- X$ decay.
The solid and the dotted line correspond to the smallest and the largest
value of ${\cal B}$ in the given parameter space, respectively. \\ \\

In Figure 2, the BR of $\bar B^0 \to \pi^- X$ is presented as a function
of $|V_{ub}|$.
We vary the value of $N_c$ and $\gamma (\equiv \phi_3)$ in a reasonable
range: from $N_c =2$ to $10$, and from $\gamma (\equiv \phi_3)=60^0$
to $110^0$.
The solid and the dotted line correspond to the smallest and the largest
value of ${\cal B}$ in the given parameter space, respectively.
The BR is an order of $10^{-4}$.
For the given $|V_{ub}|$, the BR ${\cal B}$ is estimated with a relatively
small error ($<15 \%$).
Reversely, for the given (i.e., experimentally measured) BR ${\cal B}$,
the value of $|V_{ub}|$ can be determined with a small error
($< 10 \%$).
(Of course, since in a practical experiment the BR would be measured
with some errors, $|V_{ub}|$ could be determined with larger error:
e.g., for ${\cal B} = (1.0 \pm 0.1) \times 10^{-4}$, our result expects
$|V_{ub}| = (3.7 \pm 0.47) \times 10^{-3}$.)

Using the decay process $\bar B^0 \to \pi^- X$, one may need to consider the
$B^0 - \bar B^0$ mixing effect: $\bar B^0 \to B^0 \to \pi^- X$.
The neutral $\bar B^0$ has about $18\%$ probability of decaying as the
opposite flavor $B^0$ \cite{rosner}.
It turns out \cite{klo} that even considering the effect from the
$B^0 - \bar B^0$ mixing, our result holds with resonable accuracy.

\section{Conclusion}

We have shown that among semi-inclusive charmless $B \to \pi X$ decays,
the process $\bar B^0 \to \pi^- X$ is particularly interesting and
one can determine $|V_{ub}|$ with reasonable
accuracy, by measuring the BR of $\bar B^0 \to \pi^- X$.

\end{document}